\documentclass[sigconf]{acmart}

\usepackage{subcaption}
\usepackage{graphicx}
\usepackage{lipsum}
\usepackage{makecell}
\usepackage[capitalise]{cleveref}
\usepackage{breakurl} 

\AtBeginDocument{%
  }

\copyrightyear{2024}
\acmYear{2024}
\setcopyright{rightsretained}
\acmConference[RecSys '24]{18th ACM Conference on Recommender Systems}{October 14--18, 2024}{Bari, Italy}
\acmBooktitle{18th ACM Conference on Recommender Systems (RecSys '24), October 14--18, 2024, Bari, Italy}\acmDOI{10.1145/3640457.3687164}
\acmISBN{979-8-4007-0505-2/24/10}

\begin{document}

\title{
RecSys Challenge 2024: Balancing Accuracy and Editorial Values in News Recommendations
} 

\author{Johannes Kruse}
\email{johannes.kruse@jppol.dk}
\orcid{0009-0007-5830-0611}
\affiliation{
  \institution{JP/Politikens Media Group}
  \city{Copenhagen}
  \country{Denmark}
}
\additionalaffiliation{
  \institution{Technical University of Denmark}
  \city{Kongens Lyngby}
  \country{Denmark}
}

\author{Kasper Lindskow}
\email{kasper.lindskow@jppol.dk}
\orcid{0009-0004-6412-0930}
\affiliation{
  \institution{JP/Politikens Media Group}
  \city{Copenhagen}
  \country{Denmark}
}
\additionalaffiliation{
  \institution{Copenhagen Business School}
  \city{Frederiksberg}
  \country{Denmark}
}

\author{Saikishore Kalloori}
\email{ssaikishore@ethz.ch}
\orcid{0000-0002-9669-0602}
\affiliation{
  \institution{ETH Zürich}
  \city{Zürich}
  \country{Switzerland}
}

\author{Marco Polignano}
\email{marco.polignano@uniba.it}
\orcid{0000-0002-3939-0136}
\affiliation{
  \institution{University of Bari Aldo Moro}
  \city{Bari}
  \country{Italy}
}

\author{Claudio Pomo}
\email{claudio.pomo@poliba.it}
\orcid{0000-0001-5206-3909}
\affiliation{
  \institution{Politecnico di Bari}
  \city{Bari}
  \country{Italy}
}

\author{Abhishek Srivastava}
\email{abhishek@iimv.ac.in}
\orcid{0000-0002-7113-6947}
\affiliation{
  \institution{Indian Institute of Management Visakhapatnam}
  \city{Visakhapatnam}
  \country{India}
}

\author{Anshuk Uppal}
\email{ansup@dtu.dk}
\orcid{0009-0008-2427-5857}
\affiliation{
  \institution{Technical University of Denmark}
  \city{Kongens Lyngby}
  \country{Denmark}
}

\author{Michael Riis Andersen}
\email{miri@dtu.dk}
\orcid{0000-0002-7411-5842}
\affiliation{
  \institution{Technical University of Denmark}
  \city{Kongens Lyngby}
  \country{Denmark}
}

\author{Jes Frellsen}
\email{jefr@dtu.dk}
\orcid{0000-0001-9224-1271}
\affiliation{
  \institution{Technical University of Denmark}
  \city{Kongens Lyngby}
  \country{Denmark}
}

\renewcommand{\shortauthors}{Kruse et al.}
\acmArticleType{Review}


\begin{abstract}
    The RecSys Challenge 2024 aims to advance news recommendation by addressing both the technical and normative challenges inherent in designing effective and responsible recommender systems for news publishing. 
    This paper describes the challenge, including its objectives, problem setting, and the dataset provided by the Danish news publishers Ekstra Bladet and JP/Politikens Media Group (``Ekstra Bladet'').
    The challenge explores the unique aspects of news recommendation, such as modeling user preferences based on behavior, accounting for the influence of the news agenda on user interests, and managing the rapid decay of news items. Additionally, the challenge embraces normative complexities, investigating the effects of recommender systems on news flow and their alignment with editorial values. We summarize the challenge setup, dataset characteristics, and evaluation metrics. Finally, we announce the winners and highlight their contributions. 
    The dataset is available at: \href{https://recsys.eb.dk}{https://recsys.eb.dk}.
\end{abstract}

\maketitle

\section{Introduction}
The RecSys Challenge 2024 focuses on news recommendation, addressing both the technical and normative challenges inherent in designing effective and responsible recommender systems for news publishing. The challenge emphasizes the distinctive aspects of news recommendation, including modeling user preferences based on implicit feedback, accounting for the influence of the news agenda on user interests, and managing the rapid decay of news articles. 
Furthermore, the challenge includes the evaluation of the news flow effects of different recommender systems using beyond-accuracy metrics compared to a range of traditional editorial selection methods for news. 
We include the beyond-accuracy evaluation because understanding the effects of each recommender system on the news flow is a necessary first step in identifying their potential (mis)alignment with the editorial profile of a news publisher, as well as how their introduction at scale might affect the democratic functions of news publishers in society.
These insights are essential for publishers, and we hope they will incite further work on beyond-accuracy objectives and metrics to support value alignment, which is an emerging research agenda, see, e.g., \cite{helberger2019, lu2020_editoral_values, Vrijenhoek2021_radio, Stray2023}.

\subsection{Ekstra Bladet \& JP/Politikens Media Group}
Ekstra Bladet is a Danish newspaper published by JP/Politikens Media Group in Copenhagen, Denmark. As one of Denmark’s leading news sites, it boasts close to a million daily active users and more than 20 million daily impressions. Ekstra Bladet focuses on delivering timely and engaging news content, using innovative digital strategies to enhance the user experience.
Through the  Platform Intelligence in News-Project (PIN-Project), funded by the Innovation Fund Denmark, Ekstra Bladet has collaborated with scholars from both the social and technical sciences at Danish universities.
The PIN-Project has not only helped Ekstra Bladet improve the user experience through personalization \cite{kruse23_recsys}, but it has also been recognized with several prestigious awards for excellence in digital journalism and user engagement \cite{imna2023, wan_ifra_eu_24, wan_ifra_ww_24, dair2023}. With a commitment to providing the best digital coverage of news, Ekstra Bladet and JP/Politikens Media Group continue to set the industry standard.

\section{The Challenge Task and Problem Setting}
Personalized news recommendation has become increasingly important in the digital age, where users are overwhelmed with information from various sources \cite{Wu2023}. 
The RecSys '24 Challenge seeks to advance this field, with the primary objective of ranking a set of news articles displayed during an impression from most to least likely for a user to click on.
Participants can leverage various data points, including the user's click history, session details (such as time and device used), user metadata (including gender and age), and the articles content.
This typically involves developing recommender systems that capture both the users' interests and the contextual relevance of articles, considering factors such as the articles' content, timeliness, and the users' reading preferences.
The candidate articles are ranked, and these rankings are compared to the actual clicks made by users.



\subsection{Dataset Description}
The Ekstra Bladet News Recommendation Dataset (EB-NeRD) is a large-scale Danish dataset created by Ekstra Bladet to support advancements and benchmarking in news recommendation research. The dataset comprises $1{,}103{,}602$ unique users, $125{,}541$ distinct news articles, and a total of $37{,}966{,}985$ impression logs.

\subsubsection{Dataset Construction}
\label{sec:dataset-construction}
The dataset was collected from the impression logs of active users at Ekstra Bladet\footnote{\url{https://ekstrabladet.dk}} over a 6-week period from April 27 to June 8, 2023.
Active users were defined as those who had at least 5 and at most $1{,}000$ news click records in a 3-week period from May 18 to June 8, 2023. The lower bound ensures that the users have some level of engagement, while the upper bound helps to exclude abnormal behavior and bots.
To protect user privacy, every user was delinked from the production system and securely hashed into an anonymized ID using one-time salt hashing, see, e.g., \cite{joyofcryptography}.
\looseness=-1

Each impression log lists the articles viewed by a user during a visit at a particular time, along with the articles that were clicked. 
It includes details such as whether the user was on the front page or an article page, the timestamp, time spent, scroll percentage, the device used for browsing (i.e., computer, tablet, or mobile), and the user’s subscription status.
Users who maintain an account may have additionally disclosed specific personal details like their gender, postal code, and age. 
Along with the impression logs, the dataset features all the Danish articles present with their titles, abstracts, bodies, and metadata, including categories. 
Using proprietary models, we have enriched the dataset with entities, topics, and sentiment labels.

The dataset consists of a training, validation, and test set. Each data split has three files: 1) the impression logs for the 7-day data split period, 2) the users’ click histories, i.e., 21 days of clicked articles prior to the data split’s impression logs, and 3) all the articles present in the period. 
Finally, to include beyond-accuracy computations, the hidden test set includes $200{,}000$ impressions, which have been manufactured, i.e., the articles that are in view for these impressions are from the same set of $250$ newly published articles from the start of the test set period. Note that these impressions are only taken into account for the beyond-accuracy evaluation.

\subsection{Evaluation Methodology}
We use standard metrics to evaluate the recommendations, i.e., the Area Under the Curve (AUC), Mean Reciprocal Rank (MRR), and normalized Discounted Cumulative Gain (nDCG) \cite{Wu2020MIND}. 
The AUC, which measures how well \textit{clicked} and \textit{not-clicked} items are distinguished, is defined as
\begin{equation}
    \text{AUC} = 
        \frac{
            \sum_{i \in \mathcal{P}} \sum_{j \in \mathcal{N}} \mathbb{I}\left[p_i > n_i \right]
        }
        {
            |\mathcal{P}|  |\mathcal{N}|
         },
\end{equation}
where $\mathbb{I}\left[\cdot\right]$ is an indicator function, $\mathcal{P}$ is the set of \textit{clicked} items, $\mathcal{N}$ is the set of \textit{not-clicked} items, $p_i$ is the predicted score for the $i$-th positive item, and $n_j$ is the predicted score for the $j$-th negative item. 
Next, the MRR, which evaluates the rank of the first \textit{clicked} item, is defined as
\begin{equation}
    \label{eq:mrr}
    \text{MRR} = \frac{1}{|\mathcal{Q}|} \sum_{i \in \mathcal{Q}} \frac{1}{\text{Rank}_i},
\end{equation}
where $\mathcal{Q}$ is the set of impressions, and $\text{Rank}_i$ is the ranking of the first \textit{clicked} item in the ranked list for the $i$-th impression. 
Finally, the nDCG up to position $K$, which measures the quality of ranking considering the position of \textit{clicked} items, is defined as
\begin{equation}
    \label{eq:ndcg}
    \text{nDCG@K} = \frac{\sum_{j=1}^K (2^{r_j} - 1) / \text{log}_2(1+j)  }{ \sum_{i=1}^{\mathcal{|P|}} 1/\text{log}_2(1+i) }, 
\end{equation}
where $r_j$ is the relevance score of the $j$-th item, with $r_j = 1$ for \textit{clicked} items and $r_j = 0$ for \textit{not-clicked} items. The list is ranked such that $j=1$ corresponds to the highest-ranked item and $j=K$ to the $K$-th item (denoted by $\text{@K}$).
In the competition, we compute the metrics at the impression level, meaning that each metric is computed for each impression individually, and then the average is taken across all impressions.

\subsubsection{Beyond-Accuracy Evaluation}
To provide a more comprehensive understanding of a recommender system’s output, we also consider beyond-accuracy objectives that we evaluate with a set of mature metrics, originating from information retrieval research \cite{Smyth2001_intralistdiversity, ge2010_serendipity_coverage, Kaminskas2016} and interpret in the news publishing context, as summarized in \cref{tab:ba_interpret}. 
\begin{table*}[htbp]
    \centering
    \caption{Beyond-accuracy metrics from information retrieval research tentatively interpreted in the news publishing context.}
    \label{tab:ba_interpret}
    \begin{tabular}{|l|p{5.5cm}|p{8.6cm}|}
        \hline
        & \textbf{Conceptual definition}& \textbf{Tentative interpretation} \\
        \hline
        \textbf{Diversity} 
        & The difference between the articles recommended to users.
        & Higher diversity indicates that users are being exposed to a diverse scope of articles within each recommended list.\\
        \hline
        \textbf{Novelty} 
        & The extent to which articles have been read by the greater population of users.
        & Higher novelty indicates a low popularity bias, meaning that users are being exposed to articles that the greater population of users is not usually exposed to. \\
        \hline
        \textbf{Serendipity} 
        & The degree of match between the exposed articles and the reading histories of users.
        & Higher serendipity indicates that individual users are being exposed to articles that they do not usually read. \\
        \hline
        \textbf{Coverage} 
        & The proportion of articles from the candidate list that is exposed to users.
        & Higher coverage indicates that the population of users is being exposed to a broader scope of articles. \\
        \hline
    \end{tabular}
\end{table*}
Given a candidate list $I$ of possible items to recommend and a set of users $U$, the recommender system generates a personalized recommendation list $R_u$ for each user $u \in U$, where $R_u \subseteq I$ and $|R_u| > 1$. Furthermore, each user is associated with a click history $H_u$.
Following \citet{Smyth2001_intralistdiversity}, we measure the intralist-diversity of a recommendation as the average pairwise distance between items in the list, given by
\begin{equation}
\label{eq:intralist_diversity}
    \text{Diversity}_u = \frac{\sum_{i \in R_u} \sum_{j \in R_u \backslash \{i\}} \operatorname{dist}(i, j)}{|R_u| (|R_u| - 1)},
\end{equation}
where $\operatorname{dist}(i,j)$ denotes a distance between the $i$-th and $j$-th elements.
%
For serendipity, we follow the strategy employed by \citet{lu2020_editoral_values} and aggregate the user’s click history, then compare its similarity to each ranked article, which we aggregate and average:
\begin{equation}
\label{eq:serendipity}
    \text{Serendipity}_u = \frac{\sum_{i \in R_u} \sum_{j \in H_u} \operatorname{dist}(i, j)}{|R_u| |H_u|},
\end{equation}
where $H_u$ is the click history for user $u$. For both diversity and serendipity (Eqs.~\ref{eq:intralist_diversity}--\ref{eq:serendipity}), we employ the cosine distance function, representing each article using document embeddings. 
%
We define novelty as the negative logarithm of the popularity scores of the recommended items, i.e.,
\begin{equation}
\label{eq:novelty}
    \text{Novelty}_u = -\frac{\sum_{i \in R_u} \log_2(c_i)}{|R_u|},
\end{equation}
where $c_i$ is the popularity score of the $i$-th item, calculated in the competition as the item's normalized click count.
For coverage, we define it as the fraction of unique articles that appear across all users’ recommendation lists, given by
\begin{equation}
\label{eq:coverage}
    \text{Coverage} = \frac{|\cup_{u \in U} R_u|}{|I|},
\end{equation}
where $I$ is the candidate list.
For the metrics diversity, novelty, and serendipity, we calculate the score for each user's recommendations and report the average across all users.
We also examine the distribution of recommended articles across news categories to get an interpretable overview of the content differences for the news flow.


\subsubsection{Limitations of Evaluation Framework} 
Many socially relevant effects of news recommendation can only truly be evaluated online over longer periods (e.g., the diversity in the news diet that a recommender system exposes users to or the degree of fragmentation that personalization produces across the population via multiple interactions), which is neither possible with the dataset we provide nor the metrics we include. 
Thus, the static nature of our candidate list does not fully capture the dynamics in real-world applications.
Ultimately, taking into account the normative considerations of news publishing will require formulating goals for specific beyond-accuracy objectives that recommender systems can optimize for in order to align with the specific editorial values of a news publisher.

\section{Challenge Results}
This year’s challenge saw large participation with \textit{145 teams}, \textit{11 of which were from academic institutions}, comprising \textit{202 active participants} who collectively made \textit{774 submissions}. \Cref{tab:competion_results} shows the top results.

The \textit{:D} team combines transformers \cite{transformer_2017}, gradient boosting decision trees (GBDT) \cite{friedman2001greedy}, and ensemble techniques in a three-stage recommendation pipeline. They introduce time-aware feature engineering methods and data-splitting strategies to address the temporal nature of news articles and improve the model's generalization. 
\textit{BlackPearl} uses an ensemble of GBDT and a novel Hierarchical User Interest Modeling (HUIM) approach that distinguishes between long-term invariant interests and short-term rapidly changing interests. 
\textit{Tom3TK} employs an ensemble of GBDT that integrates article timeliness features with content-based recommendations derived from historical implicit feedback. 
Finally, \textit{FeatureSalad}, the top academic team, uses a stacking ensemble approach that combines various algorithms, including GBDT and multiple neural networks. 

Although the proposed solutions perform relatively closely in terms of AUC, they perform very differently when evaluated on beyond-accuracy metrics. 
For instance, the coverage of \textit{:D}'s solution is almost double that of \textit{BlackPearl}, even though they differ by only 1 AUC point, while \textit{Tom3TK} differs by 2 AUC points but have very similar coverage. 
Moreover, each solution generates distinct news flows for users. For example, the winning solution, \textit{:D}, recommends 42\% of articles in the news category and 5\% in the entertainment category, while the runner-up, \textit{BlackPearl}, recommends 4\% in the news category and 36\% in the entertainment category.
Other solutions recommend much higher proportions of sports content (\textit{Tom3TK} and \textit{axi2017}).
These results emphasize that while solutions perform well and similarly in offline experiments, they may have different effects on the news flow. It is important for the publishers implementing these methods to investigate their effects and ensure they align with their values and meet the needs of the end-user.
\begin{table*}[htbp]
    \centering
    \caption{
        Top 5 overall teams (1--5), 
        top 2 academic teams (I--II), 
        and four baselines: 
            (a) articles with the most clicks,
            (b) articles with the highest read-time consumption,
            (c) articles with the highest in view rate, and
            (d) a random selection.
        The evaluated metrics include AUC, MRR, nDCG@5, nDCG@10, intralist diversity (DIV), serendipity (SER), novelty (NOV), coverage (COV), and distribution across news categories: crime (CR), news (NW), sport (SP), entertainment (EN), private finances (PF), music (MU), auto-generated (AG), and miscellaneous (MS).
        Coverage and distributions are measured in percentage [\%].
        The beyond-accuracy results are computed from the top 5 articles ranked out of a candidate list of 250 articles ($\mathbf{|R|=5}$).
    }
    \label{tab:competion_results}
    \begin{tabular}{cl|cccc|cccc|cccccccc}
    \hline 
    \# & Team & AUC & MRR & \makecell{\vspace{-0.3em}nDCG\\@5} & \makecell{\vspace{-0.3em}nDCG\\@10} & DIV & SER & NOV & COV & CR & NW & SP & EN & PF & MU & AG & MS 
    \\ \hline
    1 & :D         & 89.24  & 73.27	& 79.05	& 79.85 & 0.77&	0.77& 3.70& 	63.2&	17.1&	41.6&	15.3&	4.5&	18.6&	2.2&	0.0&	0.7  \\
    2 & BlackPearl & 88.15  & 71.44	& 77.44	& 78.44 & 0.69&	0.75& 4.93& 	32.0&	1.6&	4.2&	1.9&	36.4&	15.9&	33.3&	5.0&	1.7  \\
    3 & Tom3TK     & 87.07  & 70.22	& 76.26	& 77.46 & 0.71&	0.75& 4.77& 	61.2&	14.4&	10.6&	25.3&	22.3&	24.2&	2.2&	0.5&	0.4  \\
    4 & axi2017    & 86.92  & 69.84	& 76.06	& 77.21 & 0.62&	0.78& 9.93& 	40.0&	3.8&	2.5&	32.4&	12.8&	6.1&	3.7&	34.8&	4.0  \\
    5 & hrec       & 86.67  & 68.21	& 74.63	& 75.88 & 0.76&	0.80& 11.20& 	21.6&	1.9&	28.3&	5.9&	0.3&	13.4&	0.1&	50.1&	0.0  \\
    \hline\hline 
    I &FeatureS.$^\ddagger$& 85.13  & 66.58 & 73.16 & 74.66 & -&-&-&-&-&-&-&-&-&-&-&-\\
    II&DArgk       & 77.13	& 54.57	& 61.29	& 64.82 & 0.69&	0.77& 6.41& 	33.2&	7.5&	14.8&	50.5&	22.9&	0.0&	0.0&	3.2&	1.0  \\
    \hline\hline
    a & Clicks     & 59.70	& 37.74	& 42.36 & 49.65 & 0.84&	0.79& 3.07& 	2.0&	20.0&	20.0&	40.0&	0.0&	20.0&	0.0&	0.0&	0.0  \\
    b & Read-time  & 59.49	& 37.10	& 41.79	& 49.13 & 0.78&	0.78& 3.18& 	2.0&	20.0&	60.0&	0.0&	0.0&	20.0&	0.0&	0.0&	0.0  \\
    c & In view    & 54.50	& 32.39	& 36.48	& 44.74 & 0.79&	0.79& 4.63& 	2.0&	20.0&	20.0&	20.0&	0.0&	20.0&	0.0&	0.0&	20.0 \\
    d & Random     & 49.98 & 31.56	& 34.89	& 43.38 & 0.75&	0.81& 11.10& 	100.0&	9.6&	16.0&	16.8&	8.4&	3.6&	3.6&	38.0&	3.6  \\
     \hline
     \multicolumn{18}{l}{\footnotesize $^\ddagger$\textit{FeatureSalad} did not rank the beyond-accuracy impressions.}
    \end{tabular}
\end{table*}

\subsection{Ablation Study}
During the competition, it was discovered that some of the features in EB-NeRD contained future information, such as how popular the article was going to be.
However, even without these features, participants were able to engineer features that would likely not be available in an online setup. 
In response, we decided to permit the features, recognizing that imposing restrictions on specific features could be counterproductive. Additionally, enforcing such restrictions would be challenging. Instead, participants were required to describe their models and strategies in their paper submissions, including reporting results with and without the use of potentially leaked features. In \cref{tab:ablation_results}, the winning teams' results from their ablation studies can be seen. 
While each team conducted the ablation study differently, whether by focusing on specific models within the ensembles or by testing on subsets of the dataset, it is evident that features containing future information are of significant importance. Nonetheless, even with an AUC of $76.99$, \textit{:D} would be in the top $10\%$ of teams. 
\begin{table}[htbp]
    \centering
    \caption{
        AUC scores from the ablation study on data leakage for the top 5 overall teams and the top academic team: all features ($+$), without future information ($-$), and the absolute performance difference ($\Delta$).
    }
    \label{tab:ablation_results}
    \begin{tabular}{cl|ccc}
    \hline 
    \# & Team & ($+$) & ($-$) & ($\Delta$) \\
    \hline
    1 & :D          & 88.64 & 76.99 & 11.65 \\ 
    2 & BlackPearl  & 88.15	& 82.20 & 5.95  \\ 
    3 & Tom3TK      & 85.25	& 77.75 & 7.5   \\ 
    4 & axi2017     & 86.92	& 80.15 & 6.77  \\ 
    5 & hrec        & 82.94	& 80.91 & 2.03  \\ 
     \hline\hline
    I &FeatureSalad & 81.13 & 79.01 & 2.12  \\ 
     \hline		
    \end{tabular}
\end{table}

\subsection{Common Themes}
\subsubsection{Methodology}
Most solutions leverage ensemble methods, combining various models such as gradient boosting decision trees (using frameworks like LightGBM \cite{ke2017lightgbm} and CatBoost \cite{catboost_2018}) and deep learning models (including Transformer-based models \cite{transformer_2017}). 
The few teams that experimented with collaborative-filtering-based models found them less effective in the competition, likely due to the severe cold-start problem present in EB-NeRD.

\subsubsection{Feature Engineering}
The dynamic nature of news consumption and user interests is extensively explored through the use of metadata, article representations, and temporal dynamics. These efforts often focus on addressing the cold start problem and the time-sensitive relevance of news articles.


\subsubsection{Beyond-Accuracy} 
Several papers emphasize the importance of diversity and fairness in recommendations, exploring methods to balance user interests and mitigate biases. 
The motivation is to deliver a more equitable user experience rather than solely focusing on optimizing offline AUC to achieve the best-performing model.


\section{Conclusion}
In this paper, we provided a comprehensive overview of the RecSys '24 Challenge, the dataset (EB-NeRD), metrics for performance evaluation, and key points from the winning solutions and their approaches for developing news recommender systems. 
We discussed some of the issues we, as organizers, encountered during the competition, including data leakage, and how we decided to handle it.
Despite being limited, our initial analysis of beyond-accuracy metrics indicates that different recommender systems may have very different effects on the news flow. We hope to enable further discussion of the effects that a recommender system has on the news flow, as we believe this to be of great importance.

\begin{acks}
We would like to extend our gratitude to the challenge participants and the RecSys organizers for their engagement and support. We also wish to acknowledge our employers and funding bodies, including 
Ekstra Bladet, 
JP/Politikens Media Group, 
Technical University of Denmark, 
Copenhagen Business School, 
Innovation Foundation Denmark (grant number 1044-00058B), 
and the Platform Intelligence in News-Project (grant number 0175-00014B). 
The work of Marco Polignano is supported by the PNRR project FAIR - Future AI Research (PE00000013), Spoke 6 - Symbiotic AI (CUP H97G22000210007) under the NRRP MUR program funded by the NextGenerationEU.
\end{acks}

\bibliographystyle{ACM-Reference-Format}
\bibliography{main}

\end{document}